\documentclass[twocolumn,aps,amssymb,showkeys,preprintnumbers]{revtex4}

\usepackage{graphicx}
\usepackage{dcolumn}
\usepackage{bm}

\begin{document}
\preprint{MPP-2006-99}

\title{Global topological k-defects}

\author{E. Babichev}
 \email{babichev@mppmu.mpg.de}

\affiliation{
Max-Planck-Institut f\"ur Physik
(Werner-Heisenberg-Institut),
F\"ohringer Ring 6, 80805 M\"unchen, Germany}
\affiliation{
Institute for Nuclear Research of the Russian Academy of Sciences, 
60th October Anniversary prospect 7a, 117312 Moscow, Russia}

\date{\today}

\begin{abstract}
We consider global topological defects in symmetry breaking models
with a non-canonical kinetic term. Apart from a mass parameter
entering the potential, one additional dimensional parameter arises in
such models --- a ``kinetic'' mass. The properties of defects in
these models are quite different from ``standard'' global domain
walls, vortices and monopoles, if their kinetic mass scale is
smaller than their symmetry breaking scale. In particular, depending on the
concrete form of the kinetic term, the typical size of such
a defect can
be either much larger or much smaller than the size of a standard
defect with the same potential term.
The characteristic mass of a non-standard defect, 
which might have been formed during a phase transition in the early universe, 
depends on both the temperature of a phase transition and the kinetic mass. 
\end{abstract}


\keywords{topological defects, non-linear field theories}

\maketitle

\def\a{\alpha}
\def\b{\beta}
\def\e{\epsilon}
\def\o{\omega}
\def\z{\zeta}
\def\G{\Gamma}
\def\l{\lambda}

\def\pd{\partial}
\def\vp{\varphi}
\def\d{\rm{d}}

\def\be{\begin{equation}}
\def\en{\end{equation}}
\def\la{\label}

\section{Introduction}\label{SIntro}
%
Topological defects are the class of solitons which may form as a
result of symmetry-breaking phase transitions. Defects may be zero-,
one- or two-dimensional depending on the topological properties of
the minima of the effective potential. In the condensed matter
physics topological defects arise rather commonly. The well-known
examples are the flux tubes in superconductors \cite{Abr} and
vortices in superfluid helium-4.

In the cosmological context topological defects attract much
interest because they might appear in a rather natural way during
the phase transitions in the early universe. The discrete symmetry
breaking leads to the appearance of domain walls, while the breaking
of a global or a local $U(1)$-symmetry is associated with global
\cite{glstr} and local \cite{lostr} cosmic strings correspondingly.
The localized defects or monopoles arise in the gauge model
possessing $SO(3)$ symmetry which is spontaneously broken to $U(1)$
\cite{Pol,tHo}. Note that the gauge monopoles naturally arise 
in Grand Unified Theories (GUT); e.~g. in a symmetry breaking 
phase transition $G\to H\times U(1)$ where $G$ is a 
semi-simple non-Abelian group, monopoles inevitably appear.
Similar to gauge monopoles, global monopoles can arise if a global
symmetry is broken suitably.

Adding non-linear terms to the kinetic part of the Lagrangian
makes it possible for defects to arise without a symmetry-breaking
potential term \cite{Sky}.
Non-canonical kinetic terms are rather common for effective
fields theories. During the last years Lagrangians with
non-canonical fields were intensively studied in the cosmological
context. So-called $k$-fields were first introduced in the context
of inflation \cite{k-inflation} and then $k$-essence models were
suggested as solution to the cosmic coincidence problem
\cite{k-essence}. Tachyon matter \cite{tachyon} and ghost condensate
\cite{ghost} are also examples of non-canonical fields in
cosmology. A very interesting application of $k$-fields is the
explanation of dark matter by the self-gravitating coherent state of
$k$-field matter \cite{Halo}. The production of
large gravitation waves in the models of inflation with
nontrivial kinetic term was considered in \cite{MukVik}.
The effects of scalar fields with
non-canonical kinetic terms in the neighborhood of a black hole were
investigated in \cite{BH}.

In this paper we study the properties of topological defects arising
in models with non-canonical kinetic terms. We dub such defects
as $k$-defects to distinguish them from standard ones. The existence
of non-trivial configurations is ensured by the symmetry-breaking
potential term, as in the case of ``usual'' domain walls, vortices
and monopoles. However, the kinetic term is non-trivial. We choose
it to have asymptotically a standard form for small values of
gradients, and to be different from the standard one at large
gradients. 
The asymptotically canonical form at small gradients 
guarantees the standard behavior of a
system for small perturbations and is needed to avoid problems with
singularities and/or non-dynamical behavior of fields at zero
gradients. 
By contrast, the deviation of a kinetic term from a standard one 
at large gradients makes the
properties of topological defects quite different (depending on
parameters of the Lagrangian) from the standard ones because of
non-zero gradient of fields inside defects.

The paper is organized as follows. In Sec.~\ref{SModel}, we describe our model
and derive its equations of motion. The general properties of 
topological $k$-defects are studied in Sec.~\ref{SGeneral}. 
In Sec.~\ref{SConstraints} we find
constraints on the parameters of the model. Solutions for
domain wall, global vortex and global monopole are found in Sec.~V.
We summarize and discuss results and cosmological applications 
in the concluding Sec.~\ref{SDiscussion}.

\section{Model}\label{SModel}
%
We consider the following action:
\begin{equation}
  \label{act}
  S=\int {\rm d}^4 x\left[M^4 K(X/M^4) - U(f)\right],
\end{equation}
where
\begin{equation}
  \label{X}
  X=\frac 12(\pd_\mu\phi_a\pd^\mu\phi_a),
\end{equation}
$\phi_a$ is a set of scalar fields, $a=1,2,...,N$;
$f=(\phi_a\phi_a)^{1/2}$. We will mostly consider the following
potential term, which provides the symmetry breaking:
\begin{equation}
  \label{U}
  U(\phi)=\frac{\lambda}{4}(f^2-\eta^2)^{2},
\end{equation}
where $\eta$ has a dimension of mass, while $\lambda$ is a
dimensionless constant. Note that throughout this paper we use
$(-,+,+,+)$ metric signature convention. The kinetic term $K(X)$ in
(\ref{act}) is in general some non-linear function of $X$. The
action (\ref{act}) contains two mass scales: ``usual'' potential
mass parameter $\eta$ and ``kinetic'' mass $M$. It is worth to note
that the non-linear in $X$ kinetic term unavoidably leads to the new
scale in the action. In the standard case $K=-X/M^4$ and the kinetic
mass $M$ disappears from the action. For the non-trivial choices of
kinetic term the kinetic mass enters the action and changes the
properties of the resulting topological defects.

It is more convenient to make the following redefinition of
variables to the dimensionless units:
\begin{equation}
  \label{newv}
  x\to \frac{x}{M},\, \phi_a\to M\phi_a,
\end{equation}
In terms of new variables the energy density $\e$ is also dimensionless:
$\e\to M^4 \e$. 
It is easy to see that $X\to M^4 X$ and the action becomes:
\begin{equation}
  \label{S}
  S=\int{\rm d}^4 x\left[K(X) - V(f)\right],
\end{equation}
where
\begin{equation}
  \label{V}
  V(f)=\frac{\lambda}{4}(f^2-v^2)^{2},
\end{equation}
with $f\to M f$ and $v\equiv\eta/M$ is a dimensionless quantity. The
energy-momentum tensor for the action (\ref{S}) is given by:
\begin{equation}
  \label{emt}
  T_{\mu\nu} = -K_{,X}(X)\pd_\mu\phi_a\pd_\nu\phi_a + g_{\mu\nu}K(X) - g_{\mu\nu} V(f),
\end{equation}
where comma in the subscript denotes a partial derivative with
respect to $X$.
The energy density for a static configuration, which we are
interested in ($\dot\phi=0$) is:
\begin{equation}
  \label{E}
  T_{00}=- K(X) + V(f).
\end{equation}
%
%
%
Note that for static configurations $X=\vec{\pd}\phi_a\vec{\pd}\phi_a/2$
with $\vec{\pd}\equiv \{\pd/\pd x_1,\pd/\pd x_2,\pd/\pd x_3\}$.
From the variation of action (\ref{S}) with respect to $\phi$
we obtain the equations of motion:
\begin{equation}
    \label{em}
    K_{,X} \Box\phi_a + K_{,XX} X_{,\mu}\nabla^\mu\phi_a +
    \frac{dV(f)}{df}\frac{\phi_a}{f}=0,
\end{equation}
which for the static case may be recast as follows:
\begin{equation}
    \label{em-st}
    K_{,X}(\pd_i\pd_i)\phi_a
    +  K_{,XX} (\pd_i X)(\pd_i\phi_a) +  \frac{dV}{df}\frac{\phi_a}{f}=0,
\end{equation}
To search the solution describing topological defects we use the
following ansatz:
\begin{equation}
  \label{anz}
  \phi_a(x)=f(r)\frac{x_a}{r},
\end{equation}
where 
\begin{equation}
  \label{r}
  r=\left(x_a x_a\right)^{1/2},
\end{equation}
and in (\ref{r}) the summation over $a$ from $1$ to $N$ is assumed. 
The ansatz (\ref{anz}) is the same as in the ``standard'' case 
and it implies that the field configuration 
depends on $N$ spatial coordinates, $x_1$,...,$x_N$, 
and does not depend on remaining $3-N$ ones.
Substituting (\ref{anz}) into 
(\ref{em-st}) we obtain
the equations of motion in a slightly extended form:
\begin{eqnarray}
  \label{eom1}
  K_{,X}\left[f^{\prime\prime}+\frac{(N-1)f'}{r}-\frac{(N-1)f}{r^2}\right]&&\\
  + K_{,XX}\,X'_{\rm r}f' + \frac{dV}{df}&=&0,\nonumber
\end{eqnarray}
where $'\equiv d/dr$ and
$N$ is a dimension of a defect: $N=1$ for a domain wall, 
$N=2$ for a vertex of a cosmic string and 
$N=3$ for a monopole.
In the standard case, $K(X)=-X$, and Eq.~(\ref{eom1}) takes a familiar form
describing the equations of motion for ``usual'' topological defects.

In what follows we assume that the kinetic term has the standard
asymptotic behavior at small $X$. This means that in the
perturbative regime the dynamics of systems with the considered
Lagrangians is the same as those with a canonical kinetic term. This
requirement is introduced to avoid problems at $X=0$: in the case of
$X^\beta$ with $\beta<1$ there is a singularity at $X=0$; and  for
$\beta>1$ the system becomes non-dynamical at $X=0$.

For $X \gg 1$ we restrict our attention to the following
modification of the kinetic term:
\begin{equation}
  \label{pow}
  K(X)=-X^\a.
\end{equation}
Below we find the criteria for the Lagrangians to have desired
asymptotic $X\gg 1$ in the core of a topological defect. Thus we
consider the following type of kinetic terms:
\begin{eqnarray}
  \label{model}
  K(X)=\left\{
  \begin{array}{lcl}
    -X,&& X\ll 1,\\
    -X^\a, && X\gg 1.
  \end{array}
  \right.
\end{eqnarray}
Assuming that $X\gg 1$ in the core of a defect
one can easily obtain from (\ref{pow}) and (\ref{eom1})
the equation of motion inside the core of a defect:
\begin{eqnarray}
  \label{eom2}
  f^{\prime\prime}+\frac{(N-1)f'}{r}-\frac{(N-1)f}{r^2}+
  (\a-1)\left(\ln X\right)^\prime f' \nonumber\\
  - \frac{dV/df}{\a X^{\a-1}}=0
\end{eqnarray}
As a particular example of a Lagrangian having the behavior
(\ref{model}) with $\a>1$ we choose 
\begin{equation}
  \label{m1}
  K(X)=-X-X^2,
\end{equation}
and for $\a<1$ we consider:
\begin{equation}
  \label{m2}
  K(X)=-\frac{X}{1+X^{1/3}}.
\end{equation}
We imply Eqs.~(\ref{m1}) and (\ref{m2}) to be valid for positive $X$.
For negative $X$ the kinetic terms (\ref{m1}) and (\ref{m2})
should have different forms to avoid singularities in equations of motion 
and instabilities of solutions \footnote{e.~g. one can define $K(X)=-X$ for $X<0$.}.
However in this paper we consider only the static
configurations, therefore the precise form of a Lagrangian for 
negative $X$ is irrelevant for us.
%
\section{General properties}\label{SGeneral}
%
In this section we investigate the general properties of topological
defects with non-linear in $X$ kinetic term. The equations of motion
(\ref{eom1}) for arbitrary $K(X)$ are highly non-linear and cannot
be solved analytically. We restrict our attention to the study of 
topological defects arising from Lagrangians having the
asymptotic behavior (\ref{pow}) for the kinetic term. Although the
equations of motion can not be integrated even for this particular
case, some general features can be extracted without the knowledge of 
explicit solutions.

Let us start from the region close to the  $r=0$ of a topological
defect. We assume that in the core of a defect the kinetic term can
be approximated by (\ref{pow}) which means that e.g. for models
(\ref{m1}) and (\ref{m2}) one requires $X\gtrsim 1$. Otherwise we
end up with a solution which does not differ much from the standard
one. For $r\to 0$  we search a solution in the following form:
$f(r)=Ar+Br^2+Cr^3+O(r^4)$ with unknown constants $A$, $B$ and $C$
(see, e.g. \cite{Rubakov}). Substituting this expression into
(\ref{eom1}) with potential given by (\ref{V}) and setting to zero
the coefficients for powers of $r$ in the l.h.s. we find the
asymptotic expression for $f(r)$ which is valid near $r=0$:
\begin{eqnarray}
  \label{apprf0}
  f(r)&\!\!\!=& \!\!\! Ar \\
  &-&\!\!\! \left(\frac{2}{A^2 N}\right)^{\a-2}\!\! 
  \frac{\l v^2 r^3}{A\a (N+2)(N+2\a-2)} +O(r^4).\nonumber
\end{eqnarray}
Note that the standard solutions are recovered  from (\ref{apprf0})
by setting $\a=1$.
In the above expression for $f(r)$ the constant $A$ is left undetermined
and can not be fixed through this procedure. What means, in particular,
that such important characteristics as the size of a defect and its mass
are undetermined too.

It is possible, however, to estimate roughly a size of a defect [and
hence the constant $A$ in (\ref{apprf0})] without solving explicitly
the equations of motion. For this we make the change of variables in
the action (\ref{S}) as follows:
\begin{equation}
  \label{change}
  \phi=v\psi,\, x=L y,
\end{equation}
where
\begin{equation}
  \label{LS}
  L\equiv \left(\frac{2^{1/\a}}{\sqrt{2}}\right)\frac{v^{1-2/\a}}{\lambda^{1/(2\a)}}=
  \left(\frac{2^{1/\a}}{\sqrt{2}}\right) \left(\frac{\e^{1-2/\a}}{\l}\right)^{1/4},
\end{equation}
where for the sake of convenience we introduced the characteristic
energy density in the core of a defect:
\begin{equation}
  \label{ed}
  \e\equiv \l v^4.
\end{equation}
Using (\ref{LS}) the action (\ref{S}) can be rewritten as:
\begin{eqnarray}
  \label{S change}
  S&=&-2^{4(1/\a-1)}\lambda^{1-2/\a}v^{8(1-1/\a)}\times\\
  &\times&\int{\rm d}^4 y
  \left[(\nabla_y\psi_a)^{2\a} +\left((\psi_a\psi_a)^{1/2}-1\right)^2 \right].\nonumber
\end{eqnarray}
The action (\ref{S change}) does not contain any small or large
parameters (apart from irrelevant overall prefactor), which means
that in the rescaled units $\psi$ and $y$ the typical size of a
defect is of order of $1$. In the dimensionless units $x$ the size
of a defect is given roughly by (\ref{LS}). Restoring the dimensions
we obtain for the typical size of a defect:
\begin{equation}
  \label{Lph}
  L_{ph}\sim\frac{1}{\lambda^{1/(2\a)} M}\left(\frac{\eta}{M}\right)^{1-2/\a},
\end{equation}
where subscript $ph$ stands for ``physical size''.
Note, that in the standard case the scale $M$ drops out from the
expression (\ref{Lph}), as it should be and the size of a defect is
given by $L_{ph}\sim 1/(\sqrt{\lambda}\eta)$. As one can see
from (\ref{apprf0}), the solutions for any topological defect has
the asymptotic behavior, $f\propto r$. However the constant of
proportionality in the case of non-standard kinetic term is
different from the standard one. While for $K(X)=-X$ the behavior of
$f$ is approximately as:
\begin{equation}
  \label{phi-st-0}
  f_{st} \sim \sqrt{\lambda}v^2\,r,\, r\to 0,
\end{equation}
in the case (\ref{pow}) the approximation is:
\begin{equation}
  \label{phi-X2-0}
  f_{K} \sim \e^{1/2\a}\,r,\, r\to 0,
\end{equation}
with $\e$ given by (\ref{ed}). Regardless of the choice of the masses
$M$ and $\eta$, the configuration of the fields far from the center
of the topological defect is the same as a configuration of the
fields for the standard kinetic term, because nonlinear corrections
are small for $X\ll 1$ (\ref{model}).

\begin{figure*}[t]
\includegraphics[width=\textwidth]{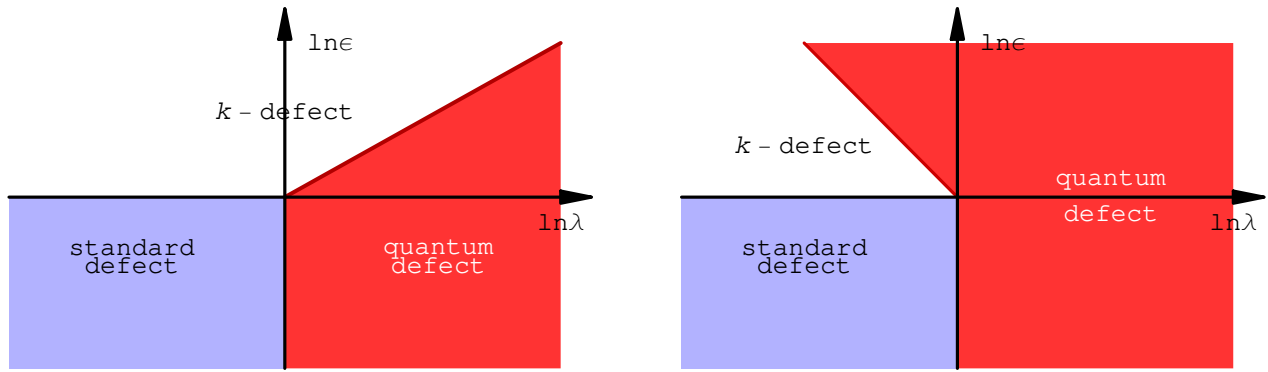}
\caption{\label{constr} Constraints on the parameters $\l$ and
$\e\equiv \l v^4$ for the model (\ref{model}) are shown in the case
$\alpha>1$ (left panel) and $\alpha<1$ (right panel). By the red
(dark-gray) color the region of parameters of the model is shown
where the classical description of topological defects fails. The
blue (light-gray) region corresponds to parameter space where the
non-linearities in $X$ are small and ``standard'' solutions are
applicable for describing topological defects.}
\end{figure*}

For a domain wall we can estimate its surface density:
\begin{equation} \label{mass}
    \sigma\equiv \int_{-\infty}^{+\infty} {\d} x\, T_{00}\sim
    \left(\frac{\e^{5-2/\a}}{\l}\right)^{1/4},
\end{equation}
where we have used (\ref{E}) and (\ref{LS}).
The ratio of the mass of a domain
$k$-wall $\sigma_k$ to the mass of a standard domain wall
$\sigma_{st}$ is given simply by the ratio of their sizes:
\begin{equation} \label{mass1}
    \frac{\sigma_k}{\sigma_{st}}\sim \frac{L_k}{L_{st}}.
\end{equation}
Restoring the dimensions we can obtain from (\ref{mass}) 
the surface energy of a domain wall in the physical units:
\begin{equation} \label{mass2}
    \sigma_{ph}\sim \sqrt{\l}\eta^3\left(\frac{\l^{1/4}\eta}{M}\right)^{2-2/\a}.
\end{equation}
Note that in the standard case, $\a=1$, the surface energy
does not depend on $M$: $\sigma_{ph}\sim \sqrt{\l}\eta^3$.

The mass per unit length of a global vortex and the total mass of a
monopole are formally divergent. However, in a cosmological context
one can define a finite mass of a monopole cutting the
integral at the position of the next monopole. The same procedure is
applicable for global cosmic strings. Masses of monopoles and cosmic
strings defined in such a way are roughly composed of two parts:
mass of a core of a defects which depends on the particular choice
of $K(X)$; and mass of a tail which depends only on the cut distance
and is the same for any $K(X)$, because for small $X$ the
non-linearities are small.
%
\section{Constraints on the parameters of the action}
\label{SConstraints}
%
Before we study particular solutions,
let us discuss  constraints on the parameters of the model.

The first constraint comes from the
condition for the hyperbolicity. One need to ensure that
the small perturbations on the background soliton solution
do not grow exponentially. For the general form
of the kinetic term $K(X)$ the hyperbolicity condition
for perturbations reads \cite{Halo,Rendall}:
\begin{equation}
  \label{hyperbolicity}
  \frac{K_{,X}(X)}{2X K_{,XX}(X)+K_{,X}(X)}>0.
\end{equation}
The stability of solutions in our case (\ref{model}) for $X\gg 1$ transforms
to the simple inequality:
\begin{equation}
  \label{hcond}
  \a>1/2.
\end{equation}
Note that the expression (\ref{hcond}) is only a necessary condition
for stability of a defect. Depending on the precise form of the
kinetic term $K(X)$ this condition can be too weak to provide the
hyperbolicity and then the more delicate consideration of inequality
(\ref{hyperbolicity}) would be required. One can check, however,
that our particular examples (\ref{m1}) and (\ref{m2}) meet the
hyperbolicity condition (\ref{hyperbolicity}) for $X>0$.

As second constraint we demand that the nonlinear part
of $K(X)$ dominates inside the core of the defect. 
Otherwise we end up
with a ``standard'' solution arising in the model of the canonical
kinetic term and a symmetry breaking potential. Thus we require
\begin{equation}
  \label{X1}
  X\gtrsim 1,
\end{equation}
which by use of Eq.~(\ref{LS}) transforms to:
\begin{equation}
  \label{Xcondg}
  \e\equiv \l v^4\gtrsim 1.
\end{equation}
And finally the third restriction takes into account the validity of
the classical description. We work in the classical description of
topological defects which is valid if  quantum effects may be
neglected. For this to be true the Compton wave length of the cube
of the with the edge $L$ should be smaller than the size of a
soliton $L$. This gives:
\be
\la{quantum}
L^4 \e \gtrsim 1,
\en
where $\e$ approximates the energy density inside the core of a defect.
Using (\ref{LS}) Eq.~(\ref{quantum}) can be rewritten as follows:
\begin{equation}
  \label{quant}
  \l \lesssim \e^{2-2/\a}.
\end{equation}
We summarize the requirement for the kinetic term to differ from
canonical one inside the core (\ref{X1}) and the validity of the
classical approximation (\ref{quantum}) for different $\a$ in
Fig.~\ref{constr}.
\begin{figure*}[t]
  \includegraphics[width=\textwidth,height=210pt]{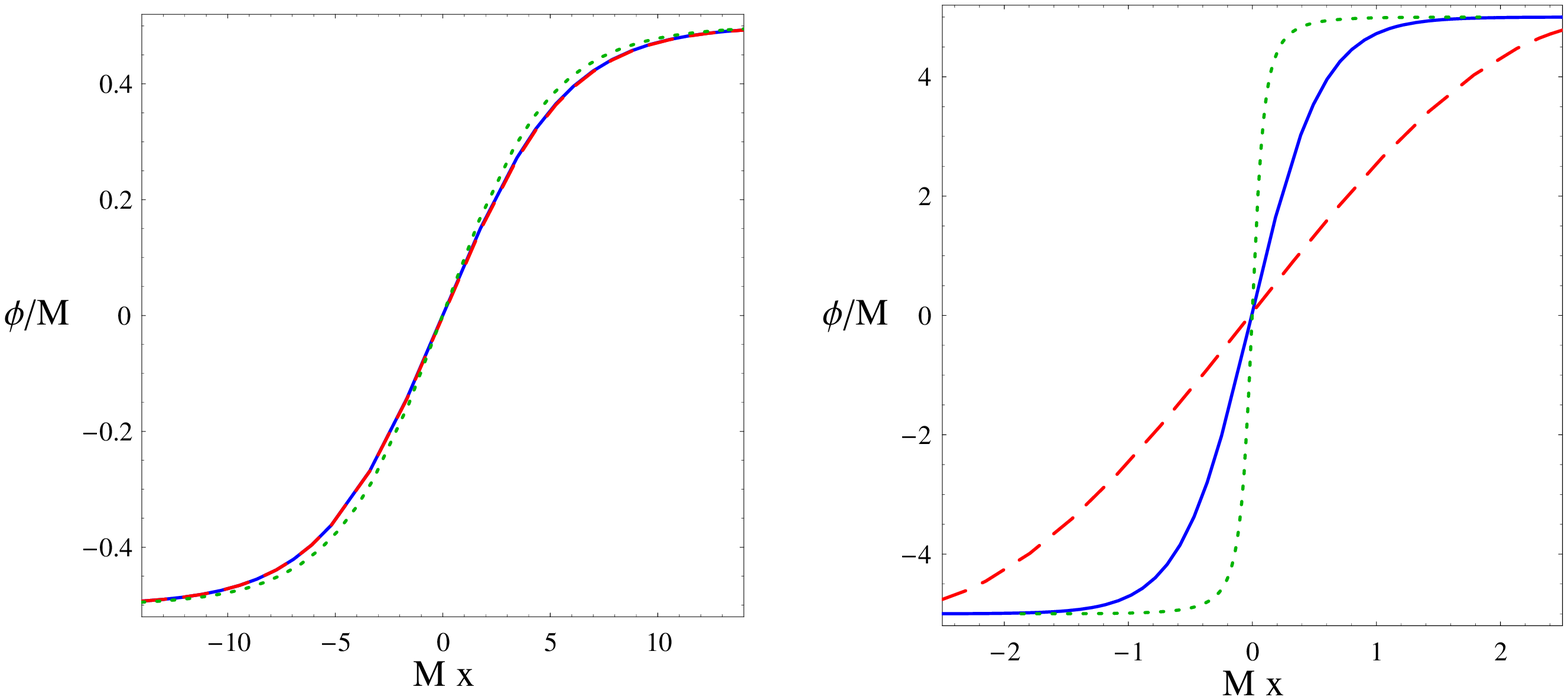}
  \caption{\label{wall} Field profiles $\phi(x)$ of domain walls
    are shown for different choices of kinetic term $K(X)$. The blue (solid)
    curve corresponds to the standard choice $K(X)=-X$, while the
    red (dashed) one accounts for the model (\ref{m1}) and the green one (dotted)
    --- for the choice of the kinetic term (\ref{m2}).
    In left panel the parameters of the models are chosen
    in such a way that the non-canonical corrections of $K(X)$
    are small: $\l=1/4$, $v=1/2$.
    In the right panel we take the parameters of the models to provide $X\gg 1$ inside
    a core of a defect: $\l=1/4$, $v=5$.}
\end{figure*}
%
\section{Solutions}\label{SSolutions}
%
We start with the simplest case of a topological defect --- a domain wall.
The solution for a domain wall arises for one real
scalar field $\phi$ depending only on one
space coordinate, say $x$.
We assume a non-linear behavior of $K(X)$ inside the topological
defect and use for definiteness the power-law dependence  (\ref{pow})
of the kinetic term for $X\gtrsim 1$.
For any $\alpha$, the equations of motion (\ref{eom2}) then read as follows:
\begin{equation}
  \label{eom w}
  (2\a-1)\phi^{\prime\prime} -\frac{1}{\a}\left(\frac{\phi^{\prime 2}}{2}\right)^{1-\a}\frac{dV}{df}=0.
\end{equation}
Eq.~(\ref{eom w}) can be integrated to give:
\begin{equation}
  \label{1int w}
  \frac{\phi'}{\sqrt{2}}=\left(\frac{V}{2\a-1}\right)^{1/(2\a)}.
\end{equation}
Let us choose the symmetry breaking potential in the form (\ref{V}).
Integrating Eq.~(\ref{1int w}) one  obtains the implicit
dependence of the function $\phi$ of $x$:
\begin{equation}
  \label{sol w}
  _2F_1\left(\frac{1}{2},\frac{1}{\a};\frac{3}{2};\frac{\phi^2}{v^2}\right)\,\phi
  = v^{2/\a}\sqrt{2}\left[\frac{\lambda}{4(2\a-1)}\right]^{1/(2\a)} x,
\end{equation}
where $ _2F_1$ is a hypergeometric function. In particular,
in the case of the standard canonical term ($\a=1$) from
(\ref{sol w}) we immediately obtain the well-known solution
for a kink:
\begin{equation}
  \label{w0}
  \phi= v\tanh\left(\sqrt{\frac{\l}{2}} v x \right).
\end{equation}
It is more interesting to find the solutions of (\ref{sol w})
in the non-standard case, $\a\neq 1$. For $\a=2$,
corresponding to the choice of the kinetic term in the
form (\ref{m1}), we find:
\begin{equation}
  \label{w1}
  f= v\sin\left[\left(\frac{\l}{3}\right)^{1/4} x \right].
\end{equation}
For $\a=2/3$ corresponding to the case (\ref{m2}) Eq.~(\ref{sol w}) gives:
\begin{equation}
  \label{w2}
  f= v\left[1+\frac{4}{(3\l)^{3/2}v^4x^2}\right]^{-1/2}.
\end{equation}
The width of defects can be estimated directly from (\ref{w0}),
(\ref{w1}) and (\ref{w2}). While in the standard case the
size of a defect is given by $L_{st}\sim \sqrt{2}/(\sqrt{\l} v)$,
for $\a=2$ and $\a=2/3$ we obtain correspondingly
\begin{equation}
  \label{La}
  L_{(2)}\sim \left(\frac{3}{\l}\right)^{1/4},\,\,\,
  L_{(2/3)}\sim \frac{2}{(3\l)^{3/4}v^2}.
\end{equation}
It is worth to note that the Eq.~(\ref{La})
is in a full agreement with (\ref{LS}).

Note that the solutions (\ref{w1}) and (\ref{w2}) are found under
the assumption $X\gg 1$ and, therefore, can not be applied for $x$
which is far from the core of the defect. Instead, the ``standard''
domain wall solution (\ref{w0}) is applicable far from the center of
a defect as $X\ll 1$ in that region. In the full region of $x$ the
solutions of the equations of motion with
non-standard kinetic terms may
be found numerically from Eq.~(\ref{eom1}). In Fig.~\ref{wall} the
standard solution (\ref{w0}) along with numerical solutions for
non-standard defects arising in models with kinetic terms (\ref{m1})
and (\ref{m2}) is shown. In the first case (the left panel of
Fig.~\ref{wall}) we choose the parameters of the models in such a
way that the non-canonical corrections of $K(X)$ are small. This
corresponds to the blue (light-gray) regions in Fig.~\ref{constr}.
In this case the solutions for non-standard topological defects are
close to the standard one as it should be. In the second case (the
right panel of Fig.~\ref{wall}) we take the parameters of the models
to provide $X\gg 1$ inside the core of the defect. One can clearly see
that $k$-defects have solutions which differ from the standard
one considerably. In particular, the size of $k$-defect with kinetic
term of the form (\ref{m1}) is larger than the size of a standard
defect, while a $k$-defect with the kinetic term (\ref{m2}) is
smaller than the standard one, in agreement with our general
conclusion (\ref{LS}).

Although the domain wall is a simplest case of topological defects,
studying its properties revealed generic
features of
$k$-defects in a simple way. While the equations of motion for
global vortices and global monopoles are more cumbersome,
$k$-vortices and $k$-monopoles in general have the same peculiar
properties as $k$-walls.

\begin{figure}[t]
\includegraphics[width=0.45\textwidth, height=200pt]{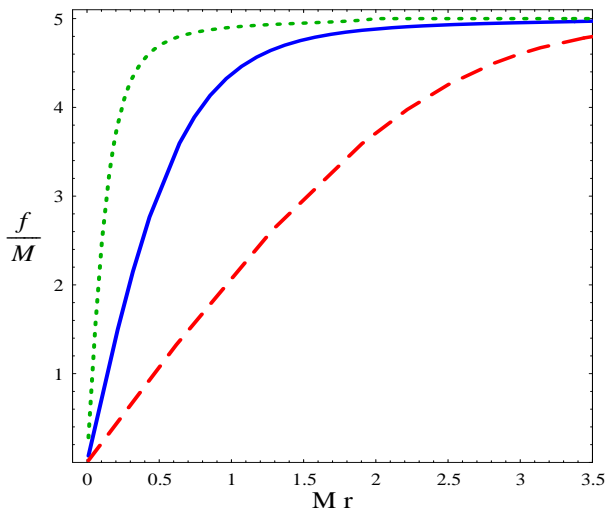}
\caption{\label{string} Field profiles $f(r)$ of cosmic strings
    are shown for different choices of the kinetic term $K(X)$. 
    The blue (solid)
    curve corresponds to the standard choice $K(X)=-X$, while the
    red (dashed) one accounts for the model (\ref{m1}) and the green one (dotted)
    --- for the choice of the kinetic term (\ref{m2}).
    The parameters of the models are taken to provide $X\gg 1$ inside
    a core of a defect: $\l=1/4$, $v=5$ (compare to the right panel of Fig.~\ref{wall}).}
\end{figure}

A global string solution arises for a scalar doublet $(\phi_1,\phi_2)$ 
with a symmetry breaking potential. In the cylindrical
coordinate system $(r,\vp,z)$ we make the following ansatz:
\begin{equation}
  \label{ans-s}
  \phi_1=f(r)\cos\vp,\,  \phi_2=f(r)\sin\vp.
\end{equation}
From Eq.~(\ref{ans-s}) we immediately find:
\begin{equation}\label{Ys}
    X=\frac12\left(f^{\prime 2}+\frac{f^2}{r^2}\right).
\end{equation}
Substituting (\ref{Ys}) into Eq.~(\ref{eom1}) and taking $\a=2$ one
can obtain the ordinary differential equation for one function
$f(r)$, which determines the profile of a defect. Unfortunately, the
solution of this equation can not be written in terms of known
functions, even for $X\gg 1$. The numerical solutions for the
function $f(r)$ describing the profile of a global k-string is
presented in Fig.~(\ref{string}) for the  standard kinetic term and
for the non-canonical choices of $K(X)$ given by (\ref{m1}) and
(\ref{m2}).

The solution for a global monopole is constructed in a similar way
as in the case of a global string. Instead of two real fields one
should take a theory with a triplet of scalar fields: $\phi_a$,
$a=1,2,3$. Taking $\a=3$ and substituting the following ansatz:
\begin{equation}
  \label{ans-m}
  \phi_a=f(r)\frac{x_a}{r}
\end{equation}
into (\ref{eom1}) one obtains an ordinary differential equation for
the function $f(r)$. The solution for $f(r)$ for different cases of
a kinetic term can be found numerically and the profile one obtains for
$f(r)$ is quite similar to that for a global string.
%
\section{Summary and Discussion}\label{SDiscussion}
%
In this work we considered the properties of topological defects
which arise in models with a symmetry breaking potential and
non-canonical kinetic term. The principal difference of the
considered models from the ``standard'' ones allowing the existence
of topological defects is the presence of non-standard kinetic part
$K(X)$, where $X$ is a canonical kinetic term (\ref{X}). The
non-linearity of the kinetic term inevitably leads to the appearance
of a new scale in the action [see Eq.~(\ref{act})], in addition to
the mass parameter in the potential term. Depending on the relation
between the kinetic and potential masses and on the precise form of
the function $K(X)$ the basic properties of topological defects can
vary drastically.

We considered general properties of defects and restrictions on
models having in mind the rather general case of a kinetic term with
the asymptotic behavior (\ref{model}). The constraints on the
parameters of the model in this case are shown in Fig.~\ref{constr}.
It turns out, that the properties of $k$-defects are quite different
from standard defects. In particular, 
the size of a defect depends both on the kinetic and a
potential masses, cf.~Eq.~(\ref{Lph}).
Depending on the parameter
$\a$ in (\ref{model}), a $k$-defect can be either smaller or greater
than the standard defect. 

As particular examples we studied in detail two concrete models
having non-canonical kinetic terms (\ref{m1}) and (\ref{m2}). The
field profiles of domain walls for different choices of $K(X)$ are
shown in Fig.~\ref{wall}: in the region of parameters where the
deviation of a kinetic term from the standard one is small, the
profiles are practically identical (left panel); if the non-linearities
in $X$ are strong inside a defect then $k$-defect has quite
different properties (right panel). The same result applies for
global cosmic strings (see Fig.~\ref{string}) and global monopoles.

In this paper we did not consider explicitly the problem of gauge models with
non-standard kinetic term. Although this problem deserves a separate 
investigation, some basic properties of local $k$-defects can be inferred, 
basing on the present work. In particular, one can estimate the
total the total mass of the $k$-monopole $M_k$ as 
the typical energy density (\ref{ed}) multiplied by the 
volume of the core of the monopole. Using (\ref{LS}) one
obtains:
\begin{equation}
  \label{MkM}
  M_k\sim M_{st} \left[\l^{1/4}\frac{\eta}{M}\right]^{6(\a-1)/\a},
\end{equation}
where $M_{st}$ is the mass of a standard monopole. 
In a similar way we find the tension of a cosmic $k$-string $\mu_k$: 
\begin{equation}
  \label{Mks}
  \mu_k\sim \mu_{st} \left[\l^{1/4}\frac{\eta}{M}\right]^{4(\a-1)/\a},
\end{equation}
where $\mu_{st}$ is the tension of a standard cosmic string.
If we consider the case of different mass scales, 
$M\ll\eta$ [compare to our requirement (\ref{Xcondg})], then 
for $\a>1$ we have $M_k \gg M_{st}$,
while for $\a<1$ we obtain $M_k \ll M_{st}$.
The same conclusion is true for tension of local cosmic $k$-strings.
We can see that, similar to global $k$-defects, 
the properties of local $k$-defects may drastically differ 
from standard local defects.
 
This result may have important consequences for
cosmological applications. In particular, standard
defects which might have been formed during phase
transitions in the early universe have a mass scale
directly connected to the temperature of a phase transition $T_c$:
$M_{st}\sim \eta \sim T_c$ for monopoles and 
$\mu_{st}\sim \eta^2 \sim T_c^2$ for cosmic strings.
In contrast, the mass scale of a resulting $k$-defect depends 
both on $T_c$ and the additional mass parameter, the kinetic mass $M$.
This effect has an interesting application 
to cosmic strings.
The new three-year WMAP
data \cite{WMAP} give an upper limit on the cosmic string tension:
$G\mu<2\times 10^{-7}$ \cite{csc1}, 
or even stronger: $G\mu<3\times 10^{-8}$ \cite{csc2}; while
the theoretical estimation of the tension 
for GUT cosmic strings gives $G\mu\sim 10^{-6}-10^{-7}$.
This means that GUT cosmic strings are almost ruled out
by the observations. However, if physics at the GUT scale involves
non-standard kinetic terms, then the GUT phase transition 
may have lead to the formation of cosmic strings with smaller 
tension, $G\mu_k \ll 10^{-6}$, thereby evading conflicts
with the present observations.

\begin{acknowledgments}
It is a pleasure to thank Professor M.~Kachelriess for useful 
discussions and critical reading of the manuscript.
This work was supported by a grant from the Alexander von Humboldt foundation.
\end{acknowledgments}


\end{document}